\documentstyle[mnextra,graphics,psfig]{mn}
\seceqnum

\def\etal{{\it et~al\ }}
\def\a370{$A370$}
\newcommand{\bq}{\begin{equation}}
\newcommand{\eq}{\end{equation}}
\newcommand{\bqr}{\begin{eqnarray}}
\newcommand{\eqr}{\end{eqnarray}}

\def\zl{z_{\rm L}}
\def\Dl{D_{\rm L}}  \def\Ds{D_{\rm S}}  \def\Dls{D_{\rm LS}}
\def\Sigcrit{\Sigma_{\rm crit}}
\def\balpha{\mbox{\boldmath$\alpha$}}
\def\bbeta{\mbox{\boldmath$\beta$}}
\def\btheta{\mbox{\boldmath$\theta$}}
\def\pderiv(#1/#2){\mathchoice{\partial#1\over\partial#2}
    {\partial#1/\partial#2} {\partial#1/\partial#2}
    {\partial#1/\partial#2}}
\def\<#1>{\langle{#1}\rangle}
\def\quo#1/#2{\mathchoice  {\hbox{$#1\over#2$}} {{#1\over#2}}
                {\scriptstyle{#1\over#2}}
                {#1\mskip-1.5mu/\mskip-1.5mu#2}}
\def\half{{\quo1/2}}

\def\Dsig{$\Delta\sigma_{mn}$}

\def\ce#1{\multicolumn{1}{c}{\hbox to 0pt{\hss#1\hss}}}

\begin{document}
\title[Mass distribution of Abell 370]{Non-parametric Reconstruction
of Cluster Mass Distribution from Strong Lensing: Modelling Abell
370\thanks{Based
on observations made with the NASA/ESA Hubble Space Telescope,
obtained from the data archive at the Space Telescope Science
Institute. STScI is operated by the Association of Universities for
Research in Astronomy, Inc. under the NASA contract NAS 5-26555.}}
 
\author[H.M. AbdelSalam, P. Saha \& L.L.R. Williams]{Hanadi
M. AbdelSalam$^1$\thanks{On leave from Physics Department, Faculty of
Science, University of Khartoum, P.O. Box 321, Khartoum,
Sudan}\thanks{e-mail: hma@astro.ox.ac.uk},
Prasenjit Saha$^1$\thanks{e-mail: saha@astro.ox.ac.uk} \& Liliya
L.R. Williams$^2$\thanks{e-mail: llrw@ast.cam.ac.uk} \\ $^1$Department of
Physics (Astrophysics), Keble Rd., OX1 3RH, Oxford\\ $^2$Institute of
Astronomy, Madingley Rd., CB3 0HA, Cambridge}
\date{\today} \maketitle

\begin{abstract}
We describe a new non-parametric technique for reconstructing the mass
distribution in galaxy clusters with strong lensing, i.e., from
multiple images of background galaxies. The observed positions and
redshifts of the images are considered as rigid constraints and
through the lens (ray-trace) equation they provide us with linear
constraint equations. These constraints confine the mass distribution
to some allowed region, which is then found by linear programming.
Within this allowed region we study in detail the mass distribution
with minimum mass-to-light variation; also some others, such as the
smoothest mass distribution.

The method is applied to the extensively studied cluster Abell 370,
which hosts a giant luminous arc and several other multiply imaged
background galaxies. Our mass maps are constrained by the observed
positions and redshifts (spectroscopic or model-inferred by previous
authors) of the giant arc and multiple image systems. The
reconstructed maps obtained for \a370 reveal a detailed mass
distribution, with substructure quite different from the light
distribution.  The method predicts the bimodal nature of the cluster
and that the projected mass distribution is indeed elongated along the
axis defined by the two dominant cD galaxies.  But the peaks in the
mass distribution appear to be offset from the centres of the cDs.

We also present an estimate for the total mass of the central region
of the cluster. This is in good agreement with previous mass
determinations. The total mass of the central region is
$M=2.0\!-\!2.7\times10^{14} M_{\odot}h_{50}^{-1}$, depending on the
solution chosen.
\end{abstract}


\section{Introduction}

Gravitational lensing operates on all scales and provides the best way
to reconstruct mass distribution, without any prior hypothesis about
the cluster dynamics or mass-to-light ratio, on large scales from
100~kpc to a few Mpc; i.e., from the innermost regions to the far
outskirts of clusters. Modelling of clusters with giant arcs directly
confirms that their innermost regions are dominated by dark matter and
thus plays an important role in probing the distribution of the dark
matter in rich clusters.

In the present paper, we describe a new method for reconstructing the
cluster Abell 370 [hereafter \a370] non-parametrically, using the
observational constraints provided by strong lensing. It is similar to
the method described in Saha \& Williams 1997 for galaxy-lenses, and
here we develop it for cluster-lenses. To our knowledge, our technique
is the first of its kind. The only ingredients needed for our
reconstruction are the positions of the multiple images, their
redshifts, the luminosity map of the lensing cluster and its redshift.
The positions of the images are taken as rigid constraints while the
luminosity distribution is a loose constraint subordinate to the
lensing data.  The model strikingly predicts the observed parameters
associated with each image, the ellipticities and orientations. The
reconstructed mass distribution compares favourably with the ROSAT/HRI
X-ray map which is also bimodal, and moreover almost all major peaks
visible in our reconstructed mass map coincide with the X-ray peaks.

Abell 370 is a very rich cluster of galaxies at redshift $z_{\rm
cl}=0.375$ and its centre appears to be dominated by two bright cD
galaxies, which are, together with their associated dark matter, {\it
mostly} responsible for the lensing. The two cD galaxies are visible
on both the optical ground-based CCD images and the HST WFC-1
image. It was just over a decade ago that \a370 was first recognised
as a lens (Lynds and Petrosian 1986).  This was confirmed by the
redshift measurement of the observed giant blue arc, $z_{\rm
arc}=0.724$ (Mellier \etal 1988, Soucail \etal 1988).
 
The giant blue luminous arc, the numerous arc(lets) and the multiple
images observed in \a370, distinguish the cluster and have made it a
target for extensive studies, observations and modelling (Soucail
\etal 1987, Narasimha \& Chitre 1988, Kovner 1989, Grossman \& Narayan
1989 and Kneib \etal 1993 [hereafter K93]).  K93, from their superb
ground-based CCD image, presented clear evidence that the giant arc
consists of at least three multiple merging images, characterising it
as a cusp-arc. They presented a fit for the cluster-lens, with the
giant arc as three merging images, assuming a bimodal mass
distribution calibrated with the giant arc's redshift and
parameterised by the ellipticities and orientations of the two
dominant cD galaxies. Such simple mass models may ignore substructures
in the cluster mass map, leading to imprecise inversion for
determining the redshifts of the background images (Kneib \etal
1996). However, their model-inferred redshifts for some of the
multiple images were subsequently confirmed by spectroscopy
(J. Bezecourt, personal communication).

The remarkable structure of the giant arc in \a370 and its eastern kink 
(see below) suggests that the source is straddling a caustic and
exhibits a higher order catastrophe than a cusp catastrophe. Careful
inspection of the HST image revealed that there are more than two breaks 
within the arc and tiny elongated bright knots or granules along the 
arc, which indicates immediately that the arc is, in fact, five merging
segments. The K93 model reproduced only the central three parts as 
multiple images while the two other segments emerge as single images
of part of the source. Smail \etal 1996, from their HST observations
of the giant arc, detected a possible bulge and faint spiral
structures visible on the eastern kink. Thus, they claim that the 
source is a late spiral, which is consistent with the spectroscopic 
identification. Another feature visible on the HST WFC-1 image is 
the radial arc $R$, which comprises two merging images across the 
inner critical curve.
   
Previous models for \a370 assume a predefined mass distribution for
the cluster, usually based on the parameters of the two dominant cD
galaxies, i.e., model \a370 as a bimodal cluster. The free parameters
used in characterising each clump are core radii, ellipticity and
orientation. Such a rigid way of modelling may put the resultant mass
distribution into a corner of the model space allowed by
observations. Thus, simple mass models based only on the observed
parameters of the two dominant galaxies may be inaccurate even though
they reproduce the multiple images correctly.

The structure of this paper is as follows. In section 2, we describe
the basic lensing equations used in modelling the cluster potential
and the bending angles. Sections 3 and 4 are devoted to the cluster
mass reconstruction, and to testing the robustness of the method,
respectively. Explicit use of Fermat's principle to reproduce the
exact location of the multiple images and to test the code is
explained in section 5, which also discusses individually all the
image systems used in our reconstruction.  A final discussion is given
in section 6.  We take $\Omega_0=1$ and $\Lambda=0$ throughout.

\section{The Method}

The main observables in any lensed system are the image positions,
relative magnifications for multiple images if any, the source
redshifts, and the lens redshift.  If the images are resolved, 
image ellipticities and orientations are also observable.  In this
section we develop a method for reconstructing the mass distribution
of the lens using constraints provided by (a)~the positions of
multiple images, (b)~image orientations and (c)~image ellipticities,
given the lens and source redshifts.  However, the reconstruction
carried out in the present paper uses only the multiple-image positions.

\subsection{Pixellated mass distributions}

For a source at unlensed angular position $\bbeta$ the time delay is
\begin{eqnarray}
T(\btheta) &=& {1+\zl\over c} \bigg({\Dl\Ds\over2\Dls}
               (\btheta-\bbeta)^2 \nonumber\\
           & & - {4G\over c^2} \int\! \Sigma(\btheta') \ln|\btheta-\btheta'|
                \,d^2\btheta' \bigg),
\end{eqnarray}
where $\Sigma(\btheta')$ is the surface mass density in the
lens plane at a position $\btheta'$. 
Since $\zl$ and $\Dl$ are fixed for a given cluster, we may as well
work with a scaled time delay
\begin{equation}
\label{arriv}
  \tau(\btheta) = \half(\btheta-\bbeta)^2 - {\Dls\over\Ds}{1\over\pi}
  \int\! \sigma(\btheta') \ln|\btheta-\btheta'| \, d^2\btheta'
\end{equation}
where
\begin{equation}
\label{sigcrit}
  \sigma(\btheta) = {\Sigma(\btheta)\over\Sigcrit}, \quad
  \Sigcrit = {c^2 \over 4\pi G}\Dl.
\end{equation}
Note that the critical density $\Sigcrit$ as defined in
(\ref{sigcrit}) is for a source at infinity; hence the factor of
$\Dls/\Ds$ in (\ref{arriv}).

We now consider a pixellated mass distribution for the cluster, with
$\sigma_{mn}$ denoting the surface density of the $mn$-th pixel in
units of $\Sigcrit$.  If $a$ is the pixel size, the total mass is
\begin{equation}
  a^2 \Sigcrit \sum_{mn} \sigma_{mn}.
\end{equation}
Defining
\begin{equation}
\label{pixelint}
   \psi_{mn}(\btheta) \equiv {1\over\pi}
   \int_{mn} \ln|\btheta-\btheta'| \,d^2\btheta'
\end{equation}
with the integral covering only the $mn$-th pixel, we can write the
scaled time delay (\ref{arriv}) as
\begin{equation}
  \tau(\btheta) = \half(\btheta-\bbeta)^2 - {\Dls\over\Ds}
  \sum_{mn} \sigma_{mn}\psi_{mn}(\btheta).
\end{equation}
Defining
\begin{equation}
  \balpha_{mn}(\btheta) \equiv \pderiv(/\btheta) \psi_{mn}(\btheta)
\end{equation}
and using Fermat's principle, i.e images are formed at the extrema of
time delay $\nabla_{\btheta}\tau(\btheta)=0$, leads to the lens
equation 
\begin{equation}
\label{trace}
  \btheta-\bbeta = {\Dls\over\Ds} \sum_{mn} \sigma_{mn}\balpha_{mn}(\btheta).
\end{equation}
Further defining
\begin{eqnarray}
\kappa_{mn}(\btheta) &\equiv& \frac{1}{2}
    \left(\pderiv(^2/\theta_x^2)+\pderiv(^2/\theta_y^2)\right)
        \psi_{mn}(\btheta), \\
\gamma_{mn}(\btheta) &\equiv& \frac{1}{2}
    \left(\pderiv(^2/\theta_x^2)-\pderiv(^2/\theta_y^2)\right)
        \psi_{mn}(\btheta), \\
\delta_{mn}(\btheta) &\equiv& \pderiv(^2/\theta_x\theta_y)
    \psi_{mn}(\btheta),
\end{eqnarray}
and
\begin{equation}
  H_{mn}(\btheta) = 
\left(
  \begin{array}{cc}
  \kappa_{mn}+\gamma_{mn} & \delta_{mn} \\
  \delta_{mn} & \kappa_{mn}-\gamma_{mn}
  \end{array}
\right)
\end{equation}
puts the inverse amplification matrix in the form
\begin{equation}
\label{Ainv}
  A^{-1}(\btheta) = 
\left(
  \begin{array}{cc}
  \tau_{xx} & \tau_{xy} \\
  \tau_{yx} & \tau_{yy}
  \end{array}
\right) ={\bf 1} - {\Dls\over\Ds} \sum_{mn} \sigma_{mn}
  H_{mn}(\btheta),
\end{equation}
where $\tau_{xx}=\tau_{\theta_x\theta_y}$ and so on.

The quantity $\psi_{mn}(\btheta)$ is the contribution of the $mn$-th
pixel to the potential at $\btheta$; similarly $\balpha_{mn}(\btheta)$
is the $mn$-th pixel's contribution to the bending angle, and the
second derivatives are contributions to terms in the inverse
amplification matrix.  

In this paper we will consider square pixels.  We have also
experimented with pixels that are Gaussian circles with overlapping
tails.  Gaussian pixels avoid discontinuities in the mass that square
pixels imply, but the difference in the final results is very small.
This is because the bending angle integrates once over the mass
distribution and the potential integrates twice, which tends to wash
out any effects of mass discontinuities at pixel boundaries.

Explicit expressions for $\psi_{mn}(\btheta)$ and its derivatives for
square pixels are given in Appendix A.

\subsection{Constraint equations and inequalities}
\label{constraints}
Since $\psi_{mn}(\btheta)$ and its derivatives are known functions,
the lens equation (\ref{trace}) and the inverse amplification
(\ref{Ainv}) are linear in $\sigma_{mn}$ and $\bbeta$, i.e. the 
unknowns that the reconstruction method must infer.  This linearity
renders multiple-image positions, orientations and ellipticities into
linear constraints on the pixellated mass distribution.

First consider multiple images.  For each image, we write the lens
equation (\ref{trace}) at the observed $\btheta$ and thus get a
two-component constraint equation.  But each source $\bbeta$
introduces two extra numbers to solve for, so multiple image systems
actually supply $\rm2(\<images> - \<sources>)$ constraints.

Orientations and ellipticities of single but distorted images can also
provide linear constraints.  Suppose we have an image at $\btheta$
observed elongated along position angle $\phi$, and consider the inverse
amplification matrix in coordinates $\theta_x',\theta_y'$, rotated by
$\phi$. We will have
\begin{equation}
  \label{Ainvrot}
  A^{-1} = 
  \left(\begin{array}{cc}
  \tau_{x'x'} & \tau_{x'y'} \\
  \tau_{y'x'} & \tau_{y'y'}
  \end{array}\right).
\end{equation}
having the same form as in equation (\ref{Ainv}) but with
$H_{mn}(\btheta)$ replaced by its rotated version:
\begin{equation}
  \left(\begin{array}{cc}
  \kappa_{mn}+c\gamma_{mn}+s\delta_{mn} &
  -s\gamma_{mn}+c\delta_{mn} \\
  -s\gamma_{mn}+c\delta_{mn} &
  \kappa_{mn}-c\gamma_{mn}-s\delta_{mn}
  \end{array}\right)
\end{equation}
where $c$ stands for $\cos2\phi$ and $s$ stands for $\sin2\phi$. Since
$\phi$ is known (because measured from the observed image), $A^{-1}$
in equation (\ref{Ainvrot}) is linear in $\sigma_{mn}$.  
Consider an image that appears more elongated than what is expected 
based on the intrinsic ellipticity distribution of galaxies, but
does not look like an edge-on spiral. With a rough estimate of the
galaxy's redshift, and assuming that galaxy's intrinsic ellipticity
is aligned with its lensing-induced elongation, we can infer that
its magnification along the $\theta_{x'}$ direction is at least 
$k$ times that along the perpendicular direction. In such a case we 
can write
\begin{equation}
  \label{ellip}
  k|\tau_{x'x'}| \leq |\tau_{y'y'}|.
\end{equation}
This becomes a linear constraint on the $\sigma_{mn}$ if we can infer
the image parity from a rough idea of where the critical curves are,
and thus remove the absolute value signs.  In cases where we can
confidently assert that (say) the magnification along $\theta_{x'}$ is
at least $k$ in absolute value, we can write
\begin{equation}
  -1/k \leq \tau_{x'x'} \leq 1/k;
\end{equation}
and here the parity doesn't matter.  Finally, for very large distortions
where it is clear that the amplification eigenvectors are along
$\theta_{x'}$ and $\theta_{y'}$ we can write
\begin{equation}
  \tau_{x'y'} = 0;
\end{equation}
here again parity doesn't matter.

Notice that the constraint (\ref{ellip}), usually with an equality sign, 
is the type of information that weak lensing observations provide. 
Because weak lensing generally produces mild ellipticity changes, many
galaxy images have to be averaged over to suppress the noise due
to intrinsic galaxy ellipticities and enhance the lensing signal. The 
same procedure can be applied in our case using average magnifications 
in (\ref{ellip}), instead of magnification of individual galaxies. 
This opens up the possibility of combining strong and weak lensing data 
in a mass reconstruction. In this paper we will explore only the 
multiple-image constraints and leave ellipticities and orientations for 
future work.  

\subsection{Producing a mass map}

The various observational constraints above, combined of course with
\begin{equation}
   \sigma_{mn} \geq 0,
\end{equation}
confine the pixellated mass distribution to some allowed region.
Since the constraints are all linear, it is straightforward to find
this allowed region by linear programming. But since the number of
pixels will in practice far exceed the number of constraints, the
allowed region will contain a vast family of mass distributions, all
consistent with the observations. To obtain mass maps, we need to add
more information.

There are several standard ways of adding and justifying the extra
information. For example, we could ask for the smoothest mass
distribution consistent with the data, or for a maximum entropy
distribution. But for this problem, a different figure of merit seems
more appropriate: since one of the aims of cluster mass reconstruction
is to test how well light traces mass, it is interesting to study the
mass distribution that follows the light as closely as the lensing
data allow.  We can think of this as a `minimum $M/L$ variation' mass
map. We have found that this criterion by itself tends to produce
artifacts on small scales, so we include a term that tends to smooth
over the mass distribution on short scales.  More precisely, our mass
maps minimise $\Delta^2$ subject to the linear lensing constraints,
where
\begin{eqnarray}
\label{chi}
\Delta^2 &=& \sum_{mn} \left(\sigma_{mn}
	-L_{mn}{\textstyle\sum_{kl}\sigma_{kl}}\right)^2 \nonumber\\
   &+& \epsilon^4 a^{-4} \sum_{mn}
       (\sigma_{m+1,n+1}+\sigma_{m-1,n-1}+ \nonumber\\
   & & \sigma_{m+1,n-1}+\sigma_{m-1,n+1}-4\sigma_{mn})^2.
\end{eqnarray}
In equation (\ref{chi}), $L_{mn}$ is the light associated with the
$mn$-th mass pixel, scaled so that $\sum_{mn}L_{mn}=1$.  The first
term in $\Delta^2$ thus tends to minimise mass-to-light
variations. The second term in $\Delta^2$ is a discrete version of
$\epsilon^4\int\!(\nabla^2\sigma)^2$; minimizing the integrated square
of second derivative is one way of smoothing.  As before, $a$ is the
pixel size, so $\epsilon$ can be interpreted as a smoothing scale.

The numerical problem we now have to solve is to find the minimum of a
quadratic function of the $\sigma_{mn}$ (i.e., $\Delta^2$) subject to
linear equality and inequality constraints involving the
$\sigma_{mn}$.  This type of problem is known as quadratic
programming. If there is a solution, it is unique and subroutines for
finding it are widely available. We used the NAG routine E04NFF.  The
technique is limited by storage rather than speed: for $N$ pixels the
storage required is $\simeq2N^2$.  So a few thousand pixels is the
current limit.

\section{The Lens Reconstruction}

\begin{figure}
\resizebox{\columnwidth}{!}{\includegraphics{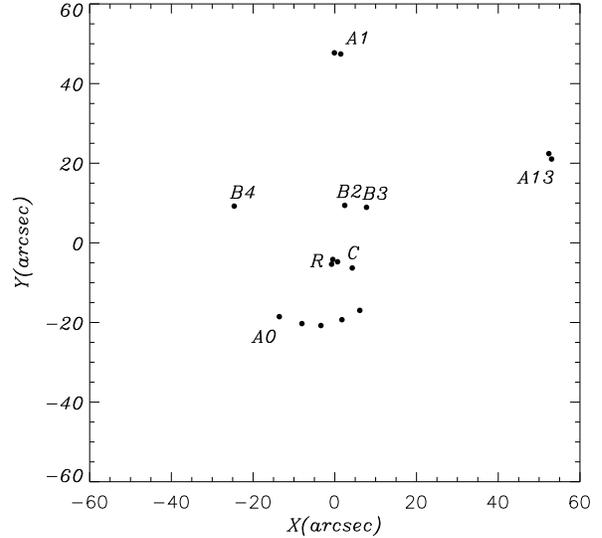}}
\caption{Multiple images systems in the cluster \a370.}
\label{images-fig}
\end{figure}

The cluster \a370 ($z_{\rm cl}=0.375$, corresponding to
$\Dl=6.1\rm\,kpc\,arcsec^{-1}$), hosts not only the first detected
giant arc but also several multiple resolved images and weakly
distorted arclets (Fort \etal 1988, K93). In this paper we will
consider only multiply imaged systems, which are in a field of
$2'\times 2'$. We take the image positions from the recent HST images
of \a370 (Smail \etal 1996) and the redshifts from K93; these are
tabulated in Table \ref{positions} and illustrated on Fig.\
\ref{images-fig}.  The critical surface mass density for sources at
infinity is $5.04\times 10^{10} h_{50}^{-1}M_{\odot}arcsec^{-2}$. 

The results we describe in this paper use pixel size $a=2.1''$;
experiments with other pixel sizes indicated little sensitivity to
$a$.
 
\subsection{Observational constraints}
\label{obs_constraints}

As explained in Section \ref{constraints}, each image contributes two
linear constraint equations but each source adds two new quantities to
solve for, and hence the number of constraints on the mass
distribution is $2\rm(\<images>-\<sources>)$.
\begin{table}
\begin{center}
\begin{tabular}{|cccrcr|}\hline 
System & Images
&\qquad& \ce{$x$(arcsec)} &\qquad& \ce{$y$(arcsec)} \\ \hline \hline 
$A0$& Giant arc &  &  \\
$(z=0.724)$     & P1 & & $-13.58$ & & $-18.53$  \\
                & P2 & & $ -8.05$ & & $-20.27$  \\
                & P3 & & $ -3.40$ & & $-20.76$  \\
                & P4 & & $  1.75$ & & $-19.30$  \\
                & P5 & & $  6.11$ & & $-16.96$  \\ \hline
$B$ \\
$(z=0.806)$     & B2 & & $  2.43$ & & $  9.41$  \\
                & B3 & & $  7.76$ & & $  8.92$  \\
                & B4 & & $-24.62$ & & $  9.22$  \\ \hline
$C$ \\
$(z=0.810)$     & C1 & & $  0.68$ & & $ -4.75$  \\
                & C2 & & $  4.27$ & & $ -6.31$  \\ \hline
$R$             & Radial arc  \\ 
$(z=1.3)$       & R1 & & $ -0.49$ & & $ -4.17$  \\
                & R2 & & $ -0.79$ & & $ -5.36$   \\\hline
$A1$ \\ 
$(z\simeq 1.2)$ & S1 & & $ -0.10$ & & $ 47.72$  \\
                & S2 & & $  1.46$ & & $ 47.43$   \\\hline
$A13$ \\
$(z=1.7\pm0.2)$ & Q1 & & $ 53.06$ & & $ 21.05$  \\
                & Q2 & & $ 52.38$ & & $ 22.41$  \\\hline
\end{tabular}
\label{positions}
\caption{Positions of the various sets of image system as taken from
the HST images.} 
\end{center}
\end{table}

\begin{itemize}
\item The giant arc $A0$ at $z_{A0}=0.724$ (Soucail \etal 1987), with
its eastern kink, was modelled by K93 as multiple merging
images of a single background source. However, the identification of
the multiple images is somewhat complex and uncertain. For example, in K93
only the central part is multiply imaged. From inspection of the HST
image it seems plausible that the giant arc is a five
image-system, and this is the interpretation we follow in this
paper. However, we found that a three-image interpretation gave a very
similar lens reconstruction. This system supplies 8 constraints. 

\item The $B2$-$B3$ and $B4$ images at $z_B=0.806$ (J. Bezecourt
private communication), which is close to the model-inferred value of
0.865 in Kneib \etal (1994 [hereafter K94]), are identified as
multiple images of a single source, and thus they provide 4 constraints.

\item The faint pair $C1$-$C2$ at $z_C=0.810$ (K93, K94)
contributes 2 constraints.

\item The recently identified radial arc $R$ at $z_R=1.3\pm0.2$,
(Smail \etal 1996) on the HST images consists of two main elongated
segments, in which a few bright granules separated by sub-arcseconds are
visible. Again we face here the uncertainty in image identification,
and we take only the position of the two main segments. Hence,
the radial arc provides us with 2 more constraints.
\end{itemize}

Most of the image systems considered above are located in the
innermost regions of the cluster and thus will constrain mostly the
region they lie in. We added a few more constraints from images
further out in the cluster, but still lying within in our field. We
included the arclets $A1$ and $A13$:
\begin{itemize}
\item The arclet $A1$ at $z_{A1}\approx 1.2$ (K94) is a
two-component distorted image, $S1$-$S2$, and adds 2 more constraints.
\item The arclet $A13$ at $z_{A13}=1.7_{-0.15}^{+0.1}$ (K94) is also a
two-component distorted image, $Q1-Q2$, and provides 2 extra
constraints. 
\end{itemize}
  
The total from all the above image systems is 20 constraints.

\begin{table}
\caption{Fiducial parameters of lensing galaxies used in our
reconstruction.  (Note that some galaxies have centroids outside our
$2'\times2'$ field, but are included here because their wings extend
into our field.)  The numbering and $B$ magnitudes are taken from
Mellier et al.\ 1989.  The two cD galaxies in this cluster are \#20 \&
\#35.}
\begin{center}
{FIDUCIAL PARAMETERS OF A370 MEMBERS}
\end{center}
\begin{center}
\begin{tabular}{|crcrcrcr|}\hline 
& \ce{MSFM No.} & \qquad\qquad
& \ce{$x{\rm (arcsec)}\pm 0.2$}  & \qquad\quad\qquad
& \ce{$y{\rm (arcsec)}\pm 0.2$}  & \qquad\quad
& \ce{$B$} \\ \hline \hline
 &  7  & & $ 23.8$ & & $ 63.5$ & & 22.22 \\
 &  8  & & $ 11.5$ & & $ 63.3$ & & 21.91 \\
 &  9  & & $ -2.4$ & & $ 62.2$ & & 21.25 \\
 & 10  & & $-31.1$ & & $ 65.1$ & & 20.13 \\
 & 12  & & $ 18.5$ & & $ 43.1$ & & 20.86 \\
 & 13  & & $ 22.8$ & & $ 53.1$ & & 21.80 \\
 & 14  & & $ 44.3$ & & $ 48.0$ & & 21.69 \\
 & 15  & & $ 56.0$ & & $ 67.8$ & & 19.68 \\
 & 16  & & $ 39.3$ & & $ 51.1$ & & 22.52 \\
 & 17  & & $-41.6$ & & $ 35.2$ & & 21.39 \\
 & 18  & & $-47.2$ & & $ 20.9$ & & 20.98 \\
 & 20  & & $ -2.8$ & & $ 25.5$ & & 20.40 \\
 & 21  & & $ -1.3$ & & $ 19.5$ & & 21.01 \\
 & 23  & & $ 56.6$ & & $ 15.6$ & & 21.07 \\
 & 28  & & $ 20.7$ & & $ -5.0$ & & 21.09 \\
 & 29  & & $-20.5$ & & $  6.7$ & & 22.82 \\
 & 30  & & $-30.0$ & & $  1.8$ & & 22.19 \\
 & 31  & & $-53.4$ & & $  5.7$ & & 21.85 \\
 & 32  & & $-55.2$ & & $ 10.8$ & & 20.44 \\
 & 34  & & $-39.9$ & & $ -4.0$ & & 21.49 \\
 & 35  & & $ -5.3$ & & $-11.3$ & & 20.93 \\
 & 36  & & $-56.4$ & & $-24.2$ & & 21.90 \\
 & 37  & & $ -8.2$ & & $-18.9$ & & 21.51 \\
 & 52  & & $ 10.4$ & & $ 25.9$ & & 21.85 \\
 & 54  & & $-32.5$ & & $ 46.5$ & & 21.64 \\
 & 56  & & $-19.9$ & & $-24.9$ & & 22.04 \\
 & 58  & & $ 15.4$ & & $-34.6$ & & 21.10 \\  \hline	
\end{tabular}
\label{fiducial}
\end{center}
\end{table}

\subsection{Pixellated light map}
\begin{figure}
\resizebox{\columnwidth}{!}{\includegraphics{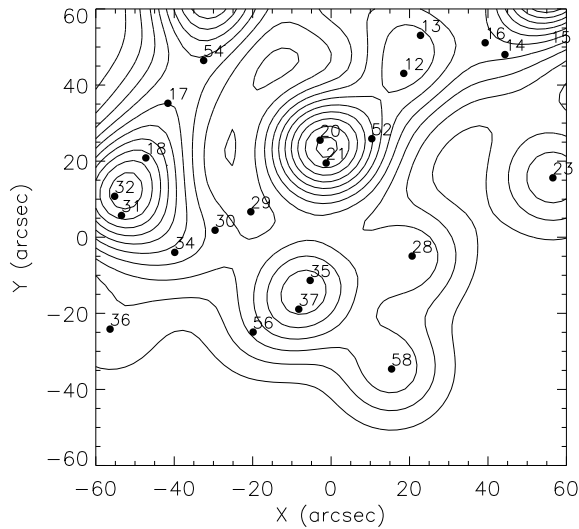}}
\caption{Luminosity map of the cluster \a370. Filled numbered circles
mark the positions of galaxies used to produce the map. The numbering
of the galaxies follows Mellier \etal 1988.}
\label{iso-fig}
\end{figure} 
The pixellated light distribution of the cluster-lens \a370 is based on 
Mellier \etal 1988. We used the apparent magnitudes of 27 individual
galaxies, including the two dominant cD galaxies [see Table
\ref{fiducial}], to obtain a smoothed luminosity map. Each of the 27 galaxies is replaced by a Gaussian light profile of dispersion $\theta_g=8''$
\bq
L_{mn}\propto\sum_{j=1}^{27} 10^{-0.4 m_j}
\exp\left(-(\btheta_{mn}-\btheta_j)^2\over 2\theta_g^2\right),
\eq
where $m_j$ is the apparent magnitude of the $j$th galaxy and
$\btheta_j$ is the location of the $j$-th
galaxy. This smoothed luminosity is plotted in Fig.~\ref{iso-fig}.

\subsection{Minimum $M/L$ variation: ML model}

\begin{figure*}
\centering
\psfig{file=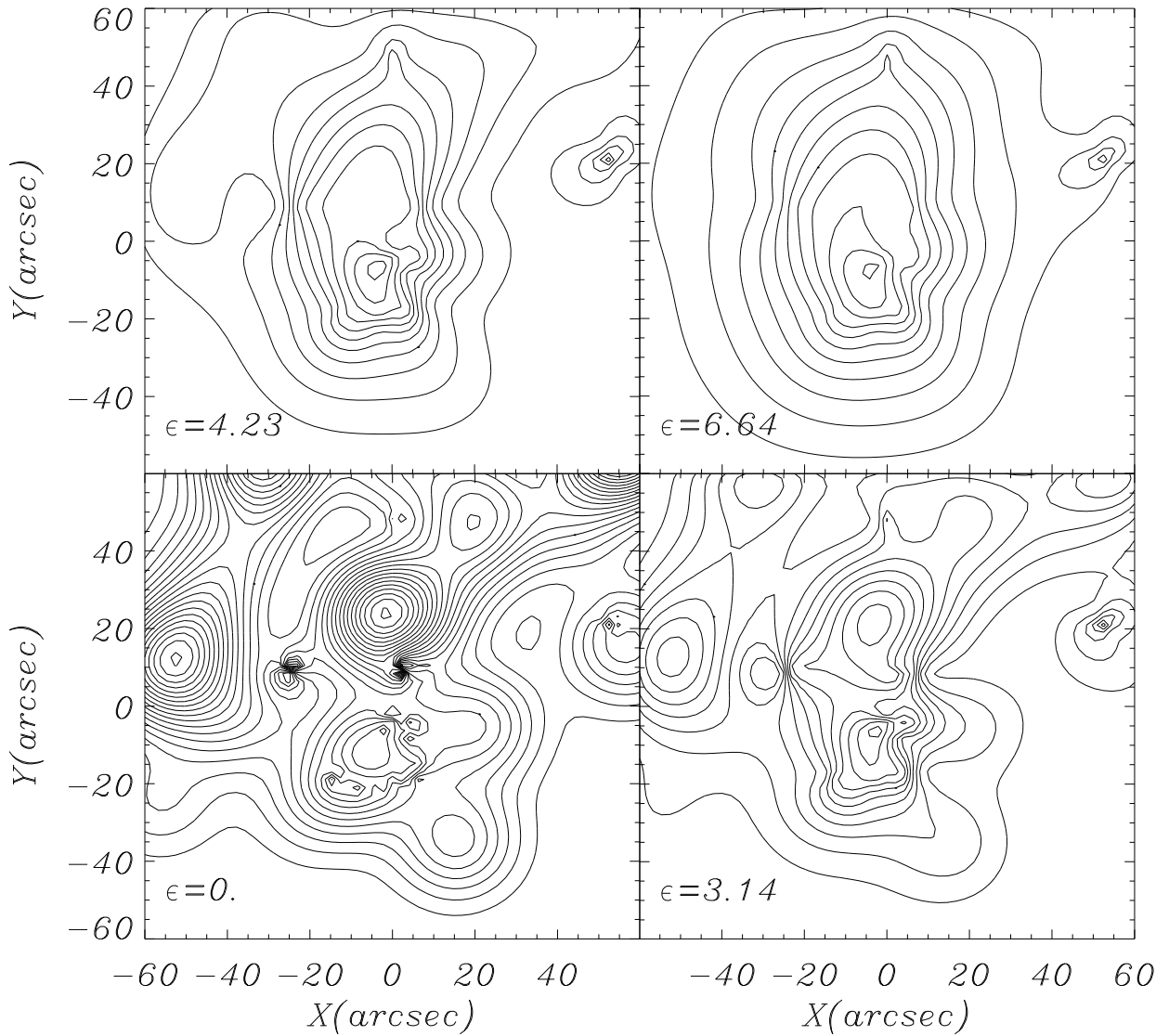,width=1.0\textwidth,angle=0}
\caption{The reconstructed mass distribution using various values for
the smoothing scale factor $\epsilon$. It is very reassuring that the
bimodality is preserved for the different values of $\epsilon$.}
\label{hess}
\end{figure*}

Now we have all the necessary ingredients to reconstruct the cluster 
mass distribution that follows the light as closely (in the
least-squares sense) as the lensing data allow.  We also have the
option of smoothing the mass distribution at various small scales
$\epsilon$ (see Eq.~[\ref{chi}]) while still satisfying the
lensing data.
\begin{figure*}
\centering
\psfig{file=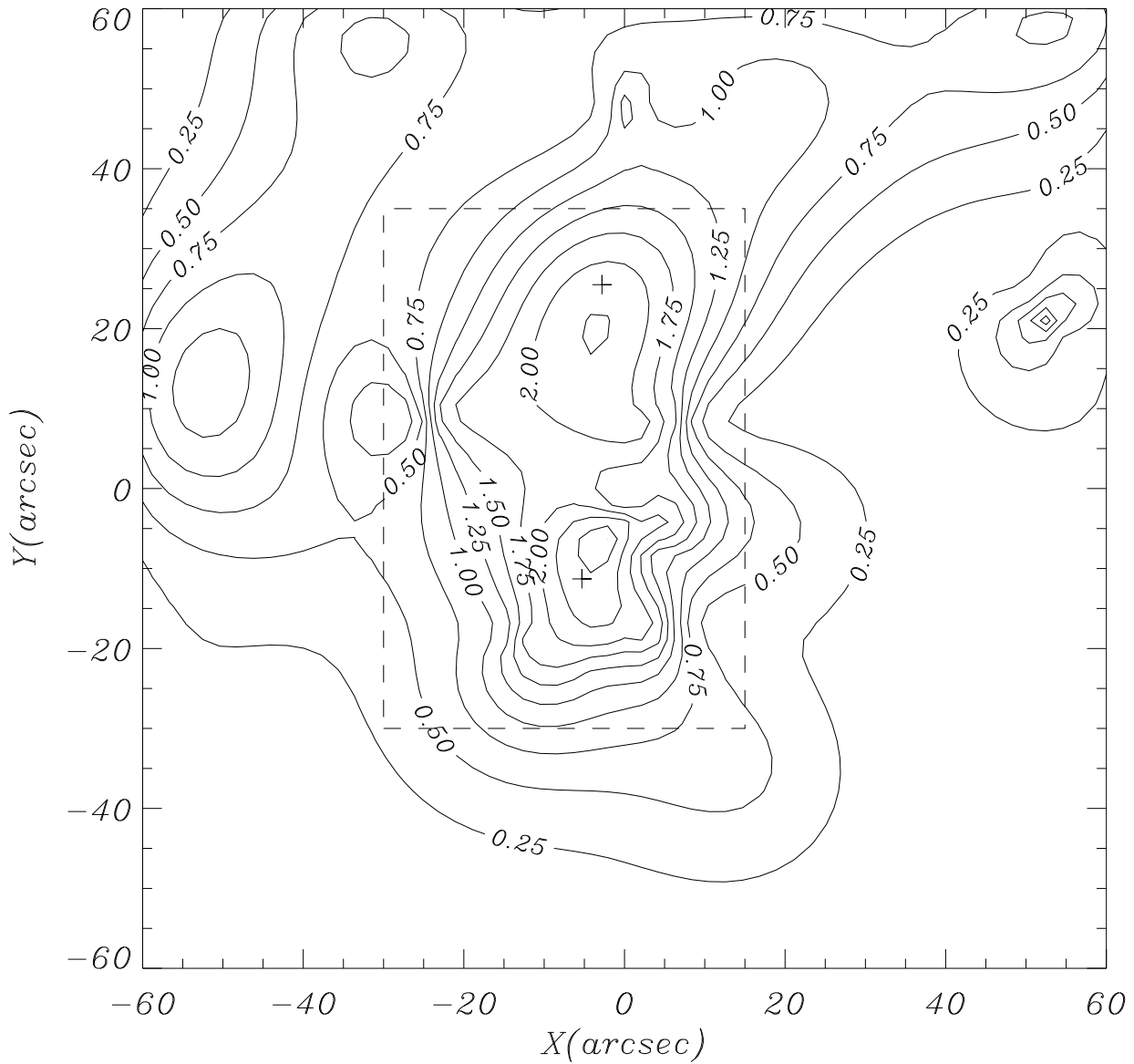,width=1.0\textwidth,angle=0}
\caption{Our best reconstructed mass distribution of the cluster
\a370. The cluster is clearly bimodal, with the two most massive
clumps in the central region close to two cD galaxies (marked with +)
in this cluster, but slightly closer together than the visible
galaxies. The dashed rectangle encloses the region where we consider
the mass reconstruction to be most robust. The contour lines are
annotated in steps of $0.25\times\Sigma_{\rm crit}$. In all the
following mass maps we adopt the same contour levels.}
\label{mass_fig}
\end{figure*}
Our reconstructed mass distribution for smoothing parameter
$\epsilon=3.5''$ is presented in Fig.~\ref{mass_fig}. Figure
\ref{hess} shows reconstructions with other values of $\epsilon$.  All
the reconstructions satisfy the lensing constraints precisely, but we
found that $\epsilon=3.5''$ is the lowest value that eliminates
evident small-scale artifacts, and in the rest of this paper we will
mainly discuss this case.

As in K93 our mass reconstruction shows bimodality in
the cluster, and this persists on mass maps for different values of
$\epsilon$---see Fig.~\ref{hess}. However, our non-parametric
reconstruction shows details of this bimodality not seen in previous
work. First, the centres of the two dominant mass clumps do not
coincide with the two dominant cD galaxies, but are slightly closer
together, roughly along the line joining the two cDs. Second, the
southern clump is much more massive than the northern clump,
although the northern cD galaxy is brighter. Third, the
mass distribution reveals extra substructure in the innermost regions
of the cluster, between the two cDs. Some of these extra substructures
are not associated with light, but are required by the lensing data,
while some are in the vicinity of cluster galaxies which did not make
it into the Mellier \etal 1989 classification scheme, and so were not
included in our isoluminosity profile, but again our model predicts
their requirement by the lensing data. Thus our non-parametric model
indicates that there is a substantial amount of dark matter in the
innermost regions of \a370, which does not closely follow the
distribution of light.  

Comparison of ROSAT/HRI X-ray map (as seen in Fort \& Mellier
1994, Mellier \etal 1994) and our reconstructed maps reveals very
similar morphologies. Reassuringly, it reveals a coincidence of the
two main central clumps that are associated with the two giant cD
galaxies. Moreover, on larger scales the X-ray map shows an elongation
towards the region of arclet $A1$ as well as secondary peaks in
north-east direction, which are well matched by peaks in our
reconstructed maps. This close similarity between the ROSAT/HRI X-ray
map for \a370 and our reconstructed maps, even on larger scales,
supports the results of our lens modelling.

\subsection{Maximally flat model: MF model}

\begin{figure}
\resizebox{\columnwidth}{!}{\includegraphics{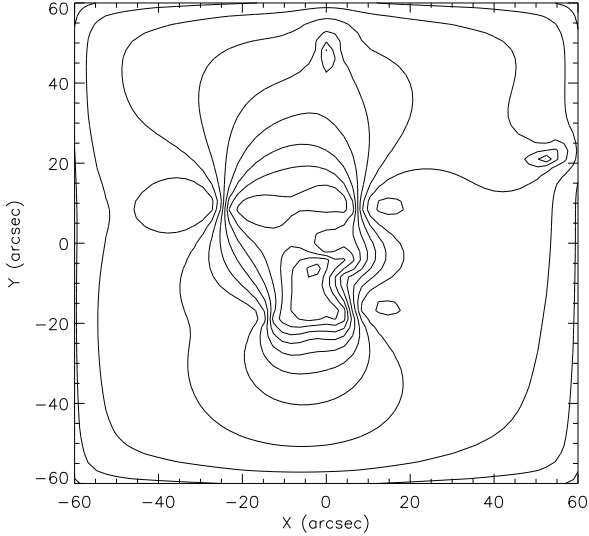}}
\caption{The reconstructed mass distribution using uniform $L_{mn}$,
which reveals the bimodal nature of the cluster.}
\label{FL}
\end{figure}

To see whether minimizing $M/L$ variation was introducing large biases
into the reconstruction, we generated another reconstruction, which we
call the MF model, with
\begin{equation}
   L_{mn} = {\rm const}.
\end{equation}
The resulting mass distribution, presented in Fig.~\ref{FL}, is the
map with the minimum dispersion in mass that still reproduces the
image positions. It is reassuring to see that in this case, we still
recover the main features of the minimum $M/L$ variation mass
map, most importantly, the bimodal nature of the cluster. Comparing
maps of Figs.\ \ref{mass_fig} and \ref{FL}, we see that even some
smaller features are reproduced in both maps.
\subsection{Uncertainties in redshifts}

Of the various multiple image systems in A370, only $A0$ and $B$ have
spectroscopic redshifts; as noted in \ref{obs_constraints}, the
redshifts of the other systems are model-inferred (K93 \& K94). To
test whether our reconstructions are sensitive to these redshifts, and
hence to model dependencies in the K93 and K94 work, we generated some
more ML maps where the model-inferred redshifts are swapped around.
The swapping $z_C\to z_R\to z_C$ and $z_C\to z_R\to z_{A13}\to z_C$
both produced maps that preserved bimodality as well as most of the extra
substructures in inner regions of the cluster; however we noticed that
more mass tends to be put in the vicinity of high redshift images.
The permutation $z_C\to z_{A13}\to z_R\to z_C$ yielded no feasible
solution; our interpretation is that if $z_{A13}$ becomes too low, it
is impossible to make the $A13$ images far enough apart without
inconsistencies with data on other images.

We thus conclude that our reconstructions are only weakly sensitive to
uncertainties in the model-inferred redshifts used.

More tests of the robustness of our reconstructions are described in the
next section.

\subsection{Mass estimates}

Our reconstructions provide mass estimates for the whole field, but as
we saw above, the mass is well constrained only in the central region,
where most of the multiple-image systems lie. Table \ref{compare}
lists our mass estimates for the central $45''\times65''$ region
(enclosed by a dashed rectangle in Fig.~\ref{mass_fig} for the ML, MF
and mean RL (see next section) cases, along with previous estimates
from the literature.  From our different reconstructions, we conclude
that 
\begin{equation}
M_{45''\times65''} = (2.3\pm0.3)h_{50}^{-1}\times 10^{14}M_\odot.
\end{equation}

\begin{table}
\caption{Previously published mass estimates of the central region of
the cluster compared with our estimates.  Our estimates refer to the
$45''\times65''$ region within the dashed rectangle in
Fig.~\ref{mass_fig}. Previous workers have usually given mass estimates
within a critical radius, so the regions considered are not precisely
equivalent.}
\begin{center}
\begin{tabular}{|ccccc|}\hline 
Model & $10^{14}h_{50}^{-1}M_{\odot}$  \\ \hline \hline 
Hammer 1987	& 2.  \\
Soucail \etal 1987  & 2.  \\
Narasimha \& Chitre 1988 & 4.2 \\
Grossman \& Narayan 1989  & 0.93 \\
Mellier \etal 1990  & 2. \\ \hline
This work: & \\ 
ML & 2.3 $\pm 0.3$ \\
MF & 2.7 \\
RL & 2.2 $\pm 0.2$ \\ \hline       
\end{tabular}
\label{compare}
\end{center}
\end{table}

The mass of the whole $2'\times 2'$
field is much less well constrained by our reconstructions. The ML
model gives
$$ M_{2'\times2'}\simeq 4.5 \times 10^{14}h_{50}^{-1}M_{\odot}, $$
while the MF model gives
$$ M_{2'\times2'}\simeq 6.6 \times 10^{14}h_{50}^{-1}M_{\odot}. $$
That the ML estimate should be lower is understandable when we think
about what the minimum $\Delta^2$ criterion [Eq.\ (\ref{chi})] does:
it follows the lensing requirements in the regions where there are
multiple images, elsewhere it tends to extrapolate using the given
$L_{mn}$. 

\section{Testing the Robustness of the method}

\begin{figure}
\resizebox{\columnwidth}{!}{\includegraphics{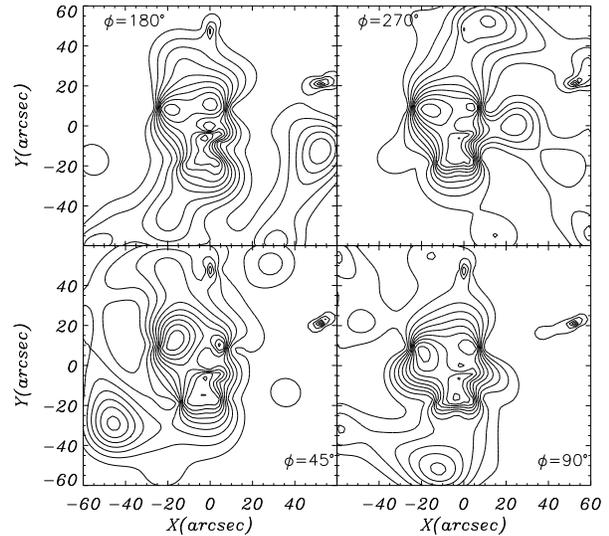}}
\caption{Some of the reconstructed mass distributions for the RL case
(see text); in each case the rotation angle $\phi$ is stated.} 
\label{rot-l}
\end{figure}

In this section we will estimate the robustness of our method. We
produce an ensemble of reconstructions, as described below, and
calculate the ensemble dispersion in mass as a function of position in
the lens plane.

\begin{figure}
\resizebox{\columnwidth}{!}{\includegraphics{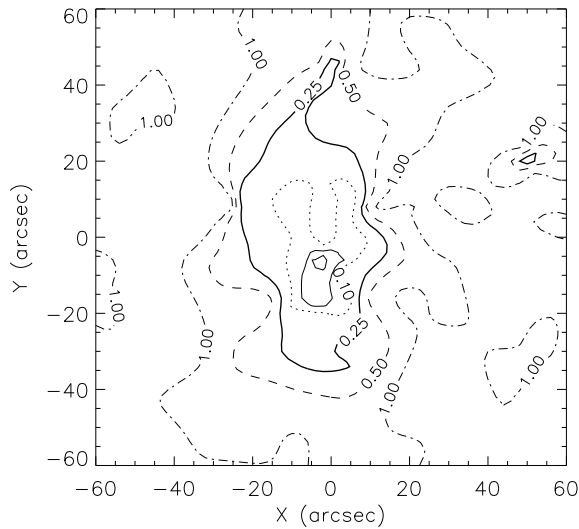}}
\caption{ Contours of \Dsig  in the lens plane. The mass
distribution is strongly constrained in the central region of the
cluster, i.e., region with high number density of images.} 
\label{rob}
\end{figure}

For each reconstruction in the ensemble, we derive $L_{mn}$ by
rotating the positions of the all the galaxy members by an angle
$\phi$ in the lens plane. Let us call these RL, rotated light,
reconstructions. Figure \ref{rot-l} shows some of these. As in the ML
and MF reconstructions, the main features of the mass map also appear
in all the versions of the RL maps. Thus we see the reconstructed mass
maps are quite robust, especially in the regions well sampled by the
observed images. We will now quantify this statement

From the RL mass distributions, we can readily calculate
pixel-by-pixel ensemble means and dispersions. Let us define
\begin{equation}
\label{ensemdisp}
   \Delta\sigma_{mn} =
   \left[{\<\sigma_{mn}^2>\over\<\sigma_{mn}>^2}-1\right]^\half
\end{equation}
which is just the ratio of ensemble dispersion to ensemble mean. It
quantifies the uncertainties in our reconstruction.  Figure
\ref{rob} shows contours of $\Delta\sigma_{mn}$ in the lens
plane. It clearly shows that the reconstructed mass distribution
is indeed very well constrained ($\Delta\sigma_{mn}\leq 0.25$) in the
inner most regions of the cluster, where almost all the multiple
images lie. This implies that the image positions indeed strongly
constrain the mass distribution enclosed by them.  Moreover, there are
two extra regions far from the centre which show some contours
revealing constrained mass distributions in their vicinity. These are
the two regions with the additional images, $A1$ and $A13$ (see table
\ref{positions}), which we added to constrain the outskirts of the
cluster. 

\begin{figure}
\resizebox{\columnwidth}{!}{\includegraphics{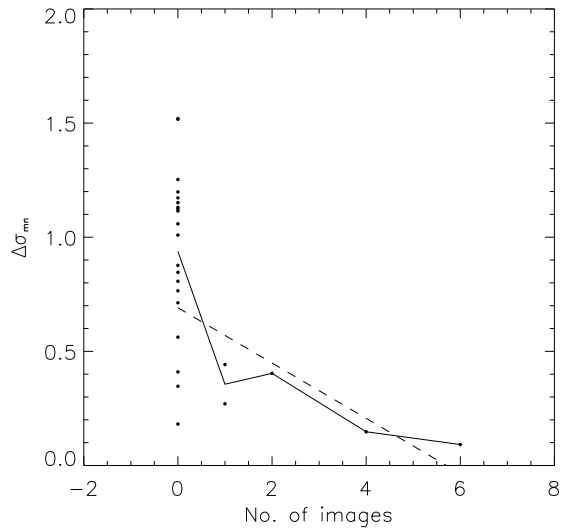}}
\caption{A plot of number of images versus \Dsig, the standard
deviation of the reconstructed mass distribution. The solid line
connects the average points of $\Delta\sigma_{mn}$ for the
corresponding numbers of images, and the dashed line is a fit.
Basically the figure shows that regions with more images are well
constrained.}
\label{robfit}
\end{figure}

Since the observed multiple images are not uniformly sprinkled over
the lens plane, the reconstructed mass distribution is not equally
well constrained over the whole lens plane; areas with more images are 
more constrained. In order to visualise this effect, in Fig.\ 
\ref{robfit} we plot the average value of $\Delta\sigma_{mn}$
computed within circles of radius $10''$ at a number of arbitrarily
placed points in the lens plane, against the number of images inclosed
by the circle.  Clearly, the robustness of the reconstructed map in
the lens plane increases with the number density of images.

\section{Discussion of the Various Image Systems}
In this section we discuss extensively all the image
systems used in our modelling and also examine their predicted shape
parameters. We start with the simplest configuration of images.

\subsection{B-System}
\begin{figure}
\resizebox{\columnwidth}{!}{\includegraphics{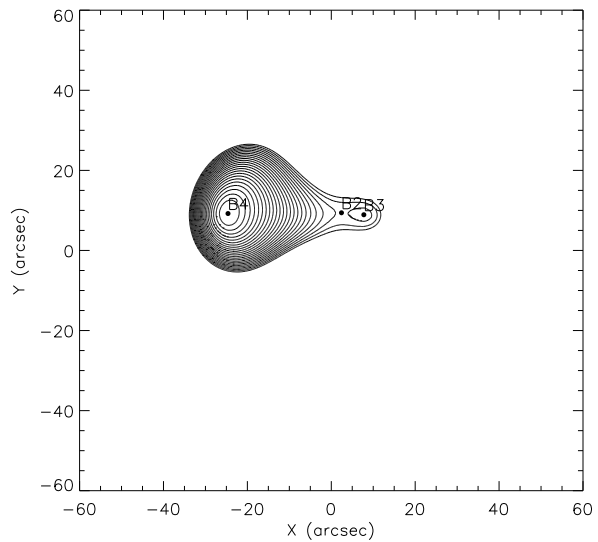}}
\caption{The arrival time contours for set $B$. Filled circles
mark the positions of the observed $B2-B3$ and $B4$. The scale
shown is in arcseconds.} 
\label{arr-b}
\end{figure}
\begin{figure}
\resizebox{\columnwidth}{!}{\includegraphics{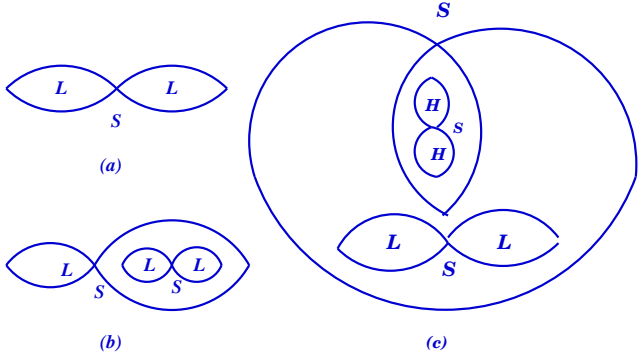}}
\caption{Inferred topological configurations, of the isochronal
contours that pass through a saddle point,for the various image sets
considered in this paper. Here (a) represents the three-image
configuration that corresponds to the $B$, $C$ $A1$ and $A13$ image
sets, (b) represents the image configuration of the giant arc $A0$ and
(c) represents the interesting seven image-configuration obtained for
the radial arc system in some reconstructions---in others the
lemniscate with two maxima and a saddle point is replaced by a single
maximum.}
\label{top}
\end{figure}

 
A good way to verify that our modelling code actually reproduces the
observed image positions is to use Fermat's principle.  In Fig.\ 
\ref{arr-b} we show a contour plot of the time delay surface for the
image set $B$ ($z_B=0.806$).  The extrema and saddle points of such a
surface mark the locations of the images produced by the model and as
expected they precisely coincide with the observed positions of $B2$,
$B3$ and $B4$. This fit indeed approves of the model.  The topography
of the arrival time surface for the $B$ image set turns out to be a
lemniscate---see Fig.\ \ref{top}a below.

\begin{figure}
\resizebox{\columnwidth}{!}{\includegraphics{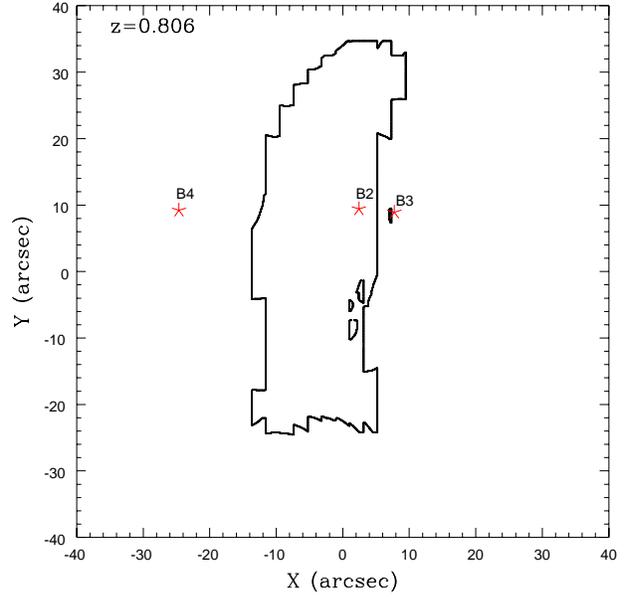}}
\caption{The critical curve at $z_B=0.806$, with the observed
positions of the images $B2$, $B3$ and $B4$ marked by asterisks.}
\label{crit-b}
\end{figure}

\begin{figure}
\resizebox{\columnwidth}{!}{\includegraphics{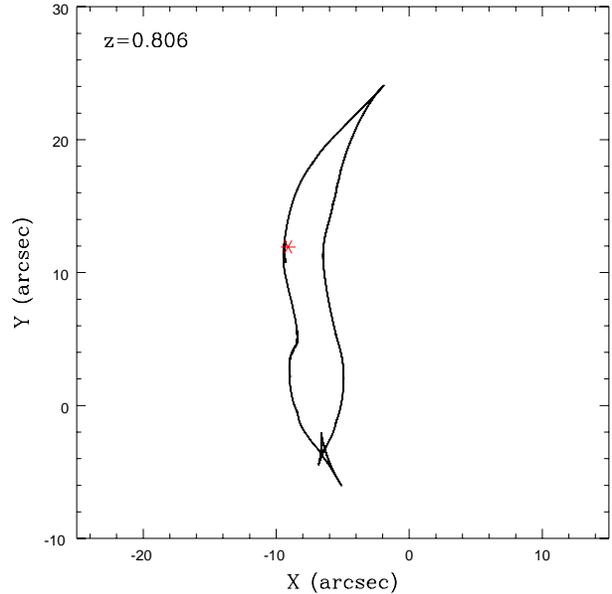}}
\caption{The caustics at $z_B=0.806$, with an asterisk marking the
inferred position of the source galaxy $B$.}
\label{caust-b}
\end{figure}

The small angular separation between the nearly merging pair $B2$-$B3$
and the fact that one is almost a mirror image of the other suggests
that a critical curve must be passing between them. We explored
critical curves with a recursive code that searches for sign changes,
and hence zeros of $\det|A^{-1}(\btheta)|$.  Figure \ref{crit-b} shows
the critical curves for the $B$ system, with the images $B2$, $B3$ and
$B4$ also indicated.
Mapping the critical curves onto the source plane, using the lens
equation, gives the caustics. Figure \ref{caust-b} shows the caustics
for the $B$ system, with the inferred position of the $B$-source also
marked.  The position of the $B$ source with respect to the
caustics show that it is experiencing a beak-to-beak fold catastrophe.

\subsection{C-System, A1 and A13 systems}
\begin{figure}
\resizebox{\columnwidth}{!}{\includegraphics{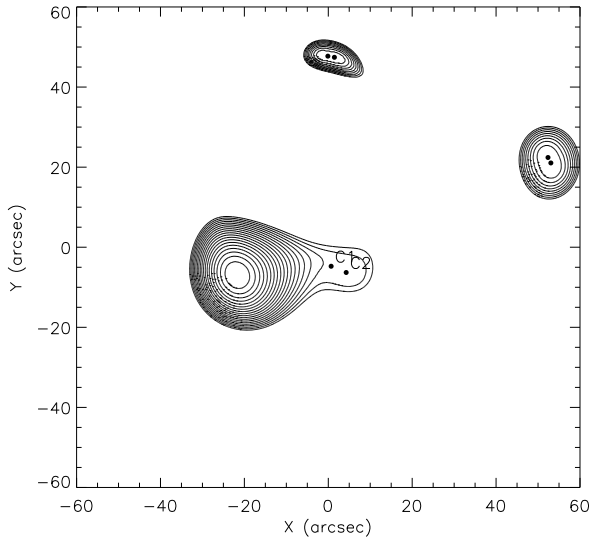}}
\caption{Time delay contours for the $C$-system and the arclets $A1$
and $A13$. The filled circles marked $C1$ and $C2$ show the
corresponding observed image positions and moreover a third image is
predicted for the same pair on the other side of the lens major
axis. The very top arrival time contour is that of the arclet $A1$ and
the one on the far right belongs to the arclet $A13$.  Though this
figure does not show such detail, these arclets are inferred to be
three-image systems too.}
\label{arr_c}
\end{figure}
The time delay surface for the $C$ system reproduce precisely the
observed position of the pair $C1$-$C2$ and also predict a third
image located on the other side of the lens major axis
(Fig.\ \ref{arr_c}). The predicted position of the third image is in the
same location as predicted before by K93 and coincides
with a faint blob in the HST image. The topological configuration of
the isochrones of the $C$ system is again a lemniscate. The caustics
at the predicted $C$ source position also exhibit a beak-to-beak fold
catastrophe.

The arclets $A1$ and $A13$ follow the same pattern of the $B$ and $C$
systems and their topological configuration correspond to lemniscates
too (see Fig.\ \ref{top}-(a)).

The lensed pair usually denoted as $D1$-$D2$, when we tried including
it, gave results similar to those for the $C1$-$C2$ pair.  However, we
have not included this system in the present paper because its
redshift is still uncertain (Kneib 1997, private communication).

\subsection{$A0$-System}

\begin{figure}
\resizebox{\columnwidth}{!}{\includegraphics{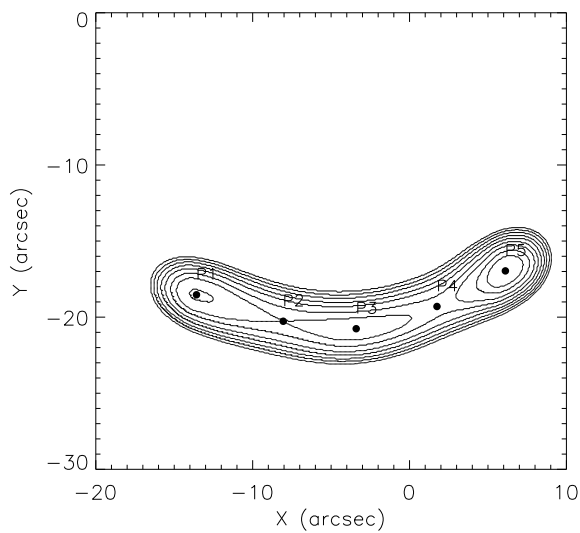}}
\caption{Time delay contours for the giant arc $A0$. Filled circles
mark the observed position of the five segments of $A0$.}
\label{arr-a}
\end{figure}

Figure \ref{arr-a} shows the time-delay contours for the giant arc
$A0$. The extrema and saddle points precisely fit the observed
positions of the five segments that constitute the giant arc. The
isochrones of this surface correspond to a lemniscate embedded in
another lemniscate (Fig.\ \ref{top}-(b)).  There is no counter-arc.

\begin{figure}
\resizebox{\columnwidth}{!}{\includegraphics{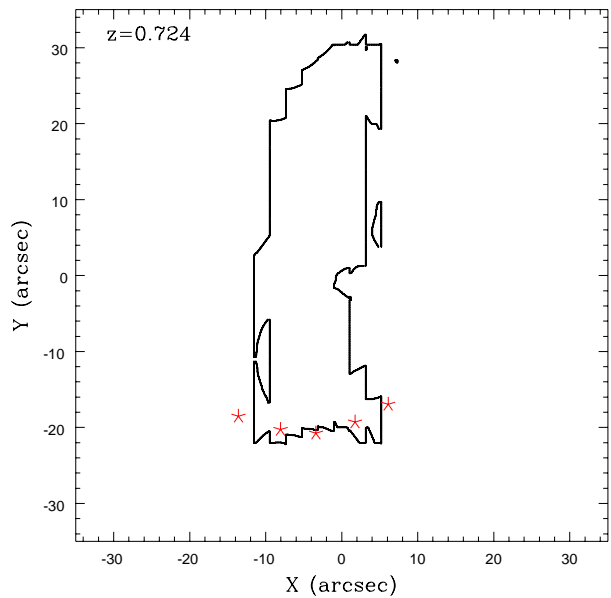}}
\caption{The critical curve at $z_{A0}=0.724$. The observed positions
of the five segments are marked with asterisks.}
\label{crit-a}
\end{figure}

\begin{figure}
\resizebox{\columnwidth}{!}{\includegraphics{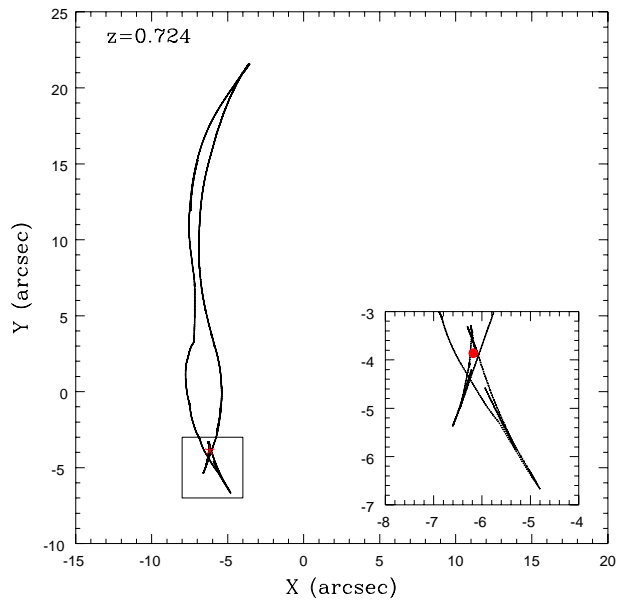}}
\caption{The caustics at $z_{A0}=0.724$.  The caustic lines in the
lower region are shown magnified in an inset. The inferred source
position, marked by a filled circle, indicates clearly that the source
is exhibiting a catastrophe which is of a higher order than a cusp.}
\label{caust-a}
\end{figure}

Figures \ref{crit-a} and \ref{caust-a} show respectively the critical
curves and caustics at $z_{A0}=0.724$.  The lower part of the latter
figure reveals a spectacular shape for the cusps. Such cusps are
called butterfly cuspoids (Berry \& Upstill 1980), and they are a
higher-order catastrophe than just a cusp. The predicted position of
the $A$ galaxy source is marked by a filled circle on Fig.\
\ref{caust-a}, which crosses the caustic at least twice. These results
for the behaviour experienced by the $A$ source galaxy indicates that
the giant arc $A0$ may be the result of higher-order catastrophe than
just the elementary cusp. If so, it would be the first giant
butterfly-cuspoid arc observed. 

\subsection{$R$-System}

Deep HST observations of clusters with spectroscopically confirmed
giant arcs has revealed lots of detailed substructures and lensed
features which had not been previously detected on ground-based CCD
images. The exceptional lensed feature visible on \a370 HST WFC-1
image (Smail \etal 1996) is the radial arc $R$, which does not show 
up even on the deep CFHT images of the cluster (K93).
Close inspection of the radial arc shows some bright blobs within the two
main elongated segments. In reconstructing the mass distribution we
used only the constraints given by the position of the two main
segments, which we call $R1$-$R2$.

\begin{figure}
\resizebox{\columnwidth}{!}{\includegraphics{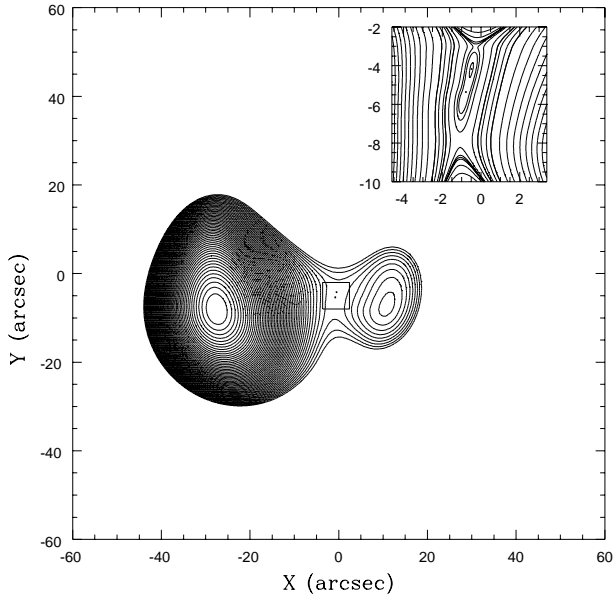}}
\caption{Contours of arrival time surface for the radial-arc. The two
filled circle, $R1-R2$, mark the position of the two main segments
considered in our mass modelling.}
\label{arr-d}
\end{figure}

>From the time-delay surfaces, we infer that the $R1$ and $R2$ are two
nearby images from a five-image or seven-image system.  Two of the
images are located on either side of the lens major axis, while the
remainder, including $R1$-$R2$, are aligned together with small
angular separation close to the southern cD galaxy.  Fig.\ \ref{arr-d}
shows the time-delay surface corresponding to Fig.\ \ref{mass_fig},
and here the radial arc consists of five aligned images.  With
Gaussian pixels, the arc was formed out of only three aligned images;
the small lemniscate was replaced by a single maximum, but otherwise
the time delay surface (and the mass distribution) was very similar.
The length of the elongated three- or five-image alignment is about
$5.2''\pm 0.2''$, which exactly fits the observed length of the radial
arc.

\begin{figure}
\resizebox{\columnwidth}{!}{\includegraphics{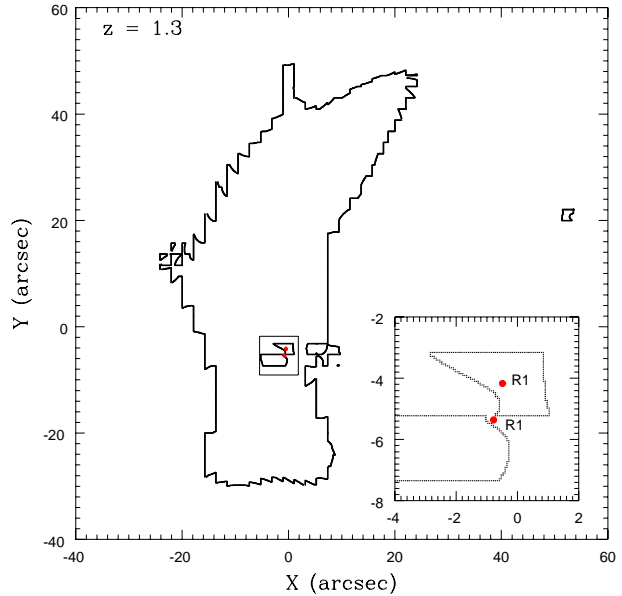}}
\caption{The critical lines at the redshift of the observed radial arc
$z_R=1.3$. Filled circles mark the position of the two segments
of the radial arc $R1-R2$. } 
\label{crit-rad}
\end{figure}

\begin{figure}
\resizebox{\columnwidth}{!}{\includegraphics{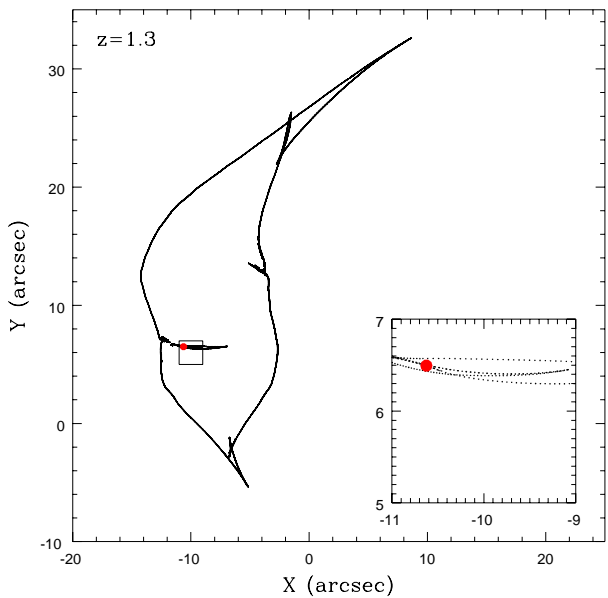}}
\caption{Caustic lines at $z_R=1.3$, with the filled circle
marking the predicted position of the $R$ source.}
\label{caust-rad}
\end{figure}

For this particular case, we recall that all our reconstructed mass
distributions predict chaotic mass fluctuations between the two cD
envelopes and detect the presence of dark sub-clumps closer to the
southern cD galaxy. These massive sub-clumps, if taken individually
would have their corresponding radial critical curves nearly
overlapping (Fig.\ \ref{crit-rad}), and hence nearly overlapping
radial caustic lines. A single extended source lying in the vicinity
of these multi-radial caustics (Fig.\ \ref{caust-rad}), will lead to
an observation of such image configuration of very close radial arcs.
  
It is of particular interest to notice the dramatic change of the
critical curves and caustics with redshift. (See for example the
sequence of Figures \ref{caust-a}, \ref{caust-b}, \ref{caust-rad}).
Notice the emergence of the radial critical curves and caustics at
high redshifts (Fig. \ref{caust-rad}). These suddenly emerged caustics
at high redshift are called lips caustics (see Schneider \etal
1992). Since our reconstruction predicts the $R$ source galaxy is
straddling these newly formed caustics, we argue that the radial arc
is a ``lips'' arc.  

\begin{table*}
\begin{minipage}{115mm}
\caption{Shape parameters of various sets of images as predicted by
the model. Almost all the parameters, i.e., orientations and
ellipticities, predicted by the model are close to the observed
ones.} 
\begin{center}
{\small Predicted shape parameters}
\end{center}
\begin{tabular}{|ccrrrr|}\hline 
Source & Images
& \multicolumn{1}{c}{$\theta$} &\multicolumn{1}{c}{$e_1$}
& \multicolumn{1}{c}{$e_2$} 
& \multicolumn{1}{c}{$\theta_{\rm obs}$}\\ \hline \hline 
$A0$& Giant arc &  &  &  \\
$(z=0.724)$& P1 & $-0.276$ & $ 0.030$ & $ 0.612$  & $-0.363 $\\
   &         P2 & $-0.231$ & $-0.034$ & $ 0.584$  & $-0.263 $\\
   &         P3 & $ 0.077$ & $ 0.025$ & $ 0.625$  & $ 0.000 $\\
   &         P4 & $ 0.435$ & $-0.084$ & $ 0.547$  & $ 0.391 $\\
   &         P5 & $ 0.705$ & $ 0.317$ & $ 0.770$  & $ 0.708 $\\ \hline
$B$&            &  &  &  & \\
$(z=0.806)$& B2 & $ 0.035$ & $-0.227$ & $ 0.400$  & $ 0.000 $\\
   &         B3 & $ 0.285$ & $ 0.628$ & $ 0.758$  & $ 0.165 $\\
   &         B4 & $-0.077$ & $ 0.518$ & $ 1.059$  & $-0.150 $\\ \hline
$C$&           &   &  & \\
$(z=0.810)$& C1 & $-0.240$ & $-0.236$ & $ 0.242$  & $-0.175$\\
   &         C2 & $-0.388$ & $ 0.144$ & $ 0.323$  & $-0.450$\\ \hline
$R$& Radial arc &  &  & \\ 
$(z=1.3)$&   R1 & $-0.191$ & $-0.634$ & $-0.020$  & $-0.192 $\\
   &         R2 & $-0.313$ & $-0.628$ & $-0.002$  & $-0.261 $ \\\hline
\end{tabular}
\end{minipage}
\label{shape}
\end{table*}

Almost all the parameters associated with the various images,
e.g. amplification, eccentricity, parity  and orientation, depend more
or less on linear combinations of the second derivatives of the 
projected potential and the images redshift. Our model predicts these
quantities for each individual image and are in good agreement with
the observations. Table~\ref{shape} lists the orientation angle
$\theta$ from the horizontal in radians and $e_1$ and $e_2$, the
two eigenvalues of the inverse amplification matrix and their ratio
gives the eccentricity. The sign of these eigenvalues specifies the
parity of the image; e.g., the image is minimum $L$ if both $e_1$ and
$e_2$ are positive, is a maximum $H$ if both are negative and a saddle
$S$ if they have opposite signs.   

\section{Conclusions}

In this paper we have developed a non-parametric technique for
reconstructing the mass distribution of galaxy clusters with strong
gravitational lensing. We divide the projected lens mass into square
pixels, and treat each pixel as an independent contributor to the
lensing potential. The observed positions of multiple images provide
linear constraints on this pixellated mass distribution. We can then
explore the mass distributions that are allowed by these constraints; a
particularly interesting model is the one that follows the light as
closely as consistency with the lensing data allow.

We applied the new method to the well known cluster Abell 370.  The
reconstructed mass maps are invariably bimodal.  The two largest mass
clumps coincide roughly with the two cD galaxies, but are closer
together than the visible cD galaxies.  Also, though the northern cD
galaxy appears to be brighter, our reconstruction reveals that the
southern mass clump is much more massive. Our reconstructions also
show various other features that may be identified with features on
the X-ray map.  In the central region, where most of the multiple
images are, the mass is well constrained; we estimate the mass of a
central $45''\times65''$ field as $2.3\pm0.3\times
10^{14}h^{-1}M_\odot$.  The mass of the whole $2'\times2'$ field is
estimated to be 4--6$\times10^{14}h^{-1}M_\odot$.

In addition, we have studied in detail the time delay surfaces,
critical curves and caustics for the various multiple image systems in
A370. In particular, we argue that the giant arc may be a five-image
system at a butterfly cuspoid catastrophe, and that the recently
discovered radial arc may be part of a seven-image system at a
lips catastrophe.
     
\section*{Acknowledgements}
We thank Jean-Paul Kneib for providing us with his 1993 data, and
James Binney for critically reading an early version and for fruitful
suggestions. Priyamavada Natarajan, Inga Schmoldt and Radek Stompor
are gratefully acknowledged for stimulating discussions. 

HMA acknowledges financial support from the Overseas Research Scheme
(ORS) and Oxford Overseas Bursary (OOB). LLRW would like to
acknowledge the support of the PPARC fellowship at the IoA, Cambridge,
UK.

\appendix
\section{}
This Appendix gives expressions for the integral over individual
pixels in equation (\ref{pixelint}), and its derivatives.  In the
following we write $x,y$ for $\theta_x,\theta_y$.

The integrals over pixels are most conveniently evaluated by first
considering the corresponding indefinite integrals and then
differencing between pixel-corner values.  The indefinite integral
correspond to equation (\ref{pixelint}) is
\begin{eqnarray}
  \tilde\psi_{mn}(x,y) &=& {1\over2\pi} \bigg(x^2\arctan{y\over x}
  + y^2\arctan{x\over y} \nonumber\\
  & & + xy\ln(x^2+y^2)-3 \bigg). \nonumber
\end{eqnarray}
Taking the pixel size as $a$ and noting that the $mn$-th pixel is
centred at $(ma,na)$, the desired definite integral is
\begin{eqnarray}
  \psi_{mn}(x,y) &=& \tilde\psi_{mn}\left(x-(m+\half)a,
                           y-(n+\half)a\right) \nonumber\\
                 &+& \tilde\psi_{mn}\left(x-(m-\half)a,
                           y-(n-\half)a\right) \nonumber\\
                 &-& \tilde\psi_{mn}\left(x-(m+\half)a,
                           y-(n-\half)a\right) \nonumber\\
                 &-& \tilde\psi_{mn}\left(x-(m-\half)a,
                           y-(n+\half)a\right). \nonumber
\end{eqnarray}
Similarly, we can compute the first derivatives of $\psi_{mn}(x,y)$ via
the indefinite integrals
\begin{eqnarray}
  \pderiv(/x)\tilde\psi_{mn}(x,y) &=& {1\over\pi}\left(
      x\arctan{y\over x} + \half y\ln(x^2+y^2) - y \right) \nonumber\\
  \pderiv(/y)\tilde\psi_{mn}(x,y) &=& {1\over\pi}\left(
      y\arctan{x\over y} + \half x\ln(x^2+y^2) - x \right), \nonumber
\end{eqnarray}
and the second derivatives via the indefinite integrals
\begin{eqnarray}
  \pderiv(^2/x^2) \tilde\psi_{mn}(x,y) &=&
     {1\over\pi}\arctan{y\over x} \nonumber\\
  \pderiv(^2/y^2) \tilde\psi_{mn}(x,y) &=&
     {1\over\pi}\arctan{x\over y} \nonumber\\
  \pderiv(^2/xy)  \tilde\psi_{mn}(x,y) &=&
     {1\over2\pi}\ln(x^2+y^2). \nonumber
\end{eqnarray}


\begin{thebibliography}{}
\bibitem[Berry 1980]{ber80}
Berry, M.V. and Upstill, C., 1980. {\it Progress in Optics}, {\bf 18}, 
257.

\bibitem[Fort 1994]{fort9}
 Bezecourt, J., 1997, {\it private communication} 

\bibitem[Fort 1988]{fort88}
 Fort, B., Prieur, J.L., Mathez, G., Mellier, Y., Soucail, G., 1988.
\newblock {\it A. \& A.}, {\bf 200}, L17. 

\bibitem[Fort 1994]{fort94}
 Fort, B. \& Mellier, Y., 1994.
\newblock {\it A. \& A. Rev.}, {\bf Volume 5, N0.4}, 637. 

\bibitem[Grossman 1989]{gross89}
 Grossman, S.A. and Narayan, R., 1989.
\newblock {\it APJ}, {\bf 344}, 637. 

\bibitem[Hammer 1989]{ham87}
 Hammer, F., 1987.
\newblock in {\it ``Third IAP Astrophys. Meeting on High Redshift
 Galaxies''}, J. Bergeron, D, Kunth, B. Rocca-Volmerange and J. Tran
 Tranh Van(eds.). Frontieres 1988. 

\bibitem[K93]{kneib93}
 Kneib, J.-P., Mellier, Y., Fort, B. and Mathez, G., 1993.
\newblock {\it A. \& A.}, {\bf 273}, 367.

\bibitem[K94]{kneib94}
 Kneib, J.-P., Mathez, G., Fort, B., Mellier, Y., Soucail, G. and
 Longaretti, P.-Y., 1994.
\newblock {\it A. \& A.}, {\bf 286}, 701.

\bibitem[Kneib \etal 1995]{kneib95}
 Kneib, J.-P., Ellis, R.S., Smail, I.R., Couch, W., Sharples, R., 1996.
\newblock {\it APJ}, {\bf 471}, 643.

\bibitem[Kovner 1989]{kovner89}
 Kovner,I.,1989.
\newblock {\it APJ}, {\bf}337, 621.

\bibitem[Lynds 1986]{lynd86}
 Lynds, R. and Petrosian, V. 1986.
\newblock {\it Bull. Am. Astronomical Soc.}, {\bf 18}, 1014.

\bibitem[Lynds 1989]{lynd89}
 Lynds, R. and Petrosian, V. 1989.
\newblock {\it APJ}, {\bf 336}, 1.

\bibitem[Mellier \etal 1988]{mill94}
 Mellier, Y., Fort,B., Bonnet, H., Kneib, J.-P., 1994, 
\newblock in ``Cosmological Aspects of X-ray clusters of galaxies''
 NATO ASI series C 441, Seitter W.C. (eds), kluwers, Dordrecht.

\bibitem[Mellier \etal 1988]{mill88}
 Mellier, Y., Soucail, G., Fort, B. and Mathez, G. 1988.
\newblock {\it  A. \& A.}, {\bf 199}, 13.

\bibitem[Mellier \etal 1990]{mill90}
 Mellier, Y., Soucail, G., Fort, B. , Le Borgne J.-F., Pello, R., 1990.
\newblock in {\it  Gravitational Lensing}, Mellier, Y., Fort, B.,
 Soucail, G. (eds), Berlin, Springer.

\bibitem[Narasimha \& Chitre 1988]{narsim88}
 Narasimha, D. \& Chitre, S.M. 1988.
\newblock {\it APJ}, {\bf 322}, 75.

\bibitem[Saha \& Williams 1997]{saha97} 
Saha, P. \& Williams, L.L.R., 1997.
\newblock {\it MNRAS}, {\bf 292}, 148.

\bibitem[Narasimha \& Chitre 1988]{narsim89}
 Schneider, P., Ehlers, J. \& Falco, E.E. 1992.
\newblock {\it Gravitational lenses}, Berlin, Springer-Verlag.

\bibitem[Smail \etal(1996)]{sm96}
 Smail, I., Dressler, A., Kneib, J-P., Ellis, R.S., Couch, W.,
 Sharples, R. and Oemler, A.Jr., 1996.
\newblock {\it APJ}, {\bf 469}, 508.

\bibitem[Soucail \etal 1987]{soc87}
 Soucail, G., Fort, B., Mellier, Y. and Picat, J.-P., 1987.
\newblock{\it  A. \& A.}, {\bf}172, L14.

\bibitem[Soucail \etal 1988]{soc88} 
Soucail, G., Mellier, Y., Fort, B. Mathez, G. and Cailloux, M., 1988.
\newblock {\it  A. \& A.}, {\bf 191}, L19


\end{thebibliography}
\end{document}